\documentstyle[aps,twocolumn,epsf]{revtex}

\begin{document}
\twocolumn[
\hsize\textwidth\columnwidth\hsize\csname @twocolumnfalse\endcsname
\title
{
Thermal and Dynamical Properties of the Two-band Hubbard Model\\
 Compared with FeSi
}

\author
{
Kentaro Urasaki$^\dagger$ and Tetsuro Saso$^{\ddagger}$
}

\address
{
Physics Department, Saitama University Urawa, Saitama 338-0825, Japan
}
\date{\today}

\maketitle

\begin{abstract}
We study the two-band Hubbard model introduced by Fu and Doniach
as a model for FeSi which is suggested to be a Kondo insulator.
Using the self-consistent second-order perturbation theory combined with the 
local approximation which becomes exact in the limit of infinite 
dimensions,
we calculate the specific heat, the spin susceptibility 
and the dynamical conductivity and point out 
that the reduction of the energy gap due to 
correlation is not significant in contrast to
the previous calculation.
It is also demonstrated that the gap at low temperatures in the optical 
conductivity
is filled up at a rather low temperature than the gap size, which is
consistent with the experiment.
\end{abstract}

]

\narrowtext

\section{Introduction}
For long time, FeSi has attracted interests of many researchers.
At low temperatures, FeSi is a nonmagnetic semiconductor with a narrow band 
gap of about 700K, and becomes magnetic and metallic when the temperature 
rises.
Jaccarino, et al.\cite{Jaccarino67} reported an anomalous decrease of 
magnetic susceptibility
at low temperatures, which could not be explained from conventional band
theories,
unless the vanishingly small band width was assumed.
Moreover, it was found by experiments that there was no magnetic ordering at 
low temperatures.\cite{Watanabe63}
Takahashi and Moriya\cite{Takahashi79} applied the spin-fluctuation theory 
to the nearly ferromagnetic semiconductor model, and
explained the temperature variation of the specific heat and the
susceptibility.
Their idea of thermally induced local moment was confirmed later by the neutron
scattering experiment.\cite{Shirane87}

On the other hand, several materials with strongly correlated f-band were 
found to
exhibit an energy gap and become insulators at low temperatures.   They are 
called Kondo insulators and have been studied intensively.
Typical examples are YbB$_{12}$,\cite{YbB12} 
Ce$_3$Bi$_4$Pt$_3$\cite{Ce3Bi4Pt3} and many others.
Aeppli and Fisk\cite{Aeppli92} suggested that FeSi may be viewed as a Kondo 
insulator in
d-electron systems, a band insulator with moderate electron correlation.
Schlesinger, et al.\cite{Schlesinger93} measured the optical conductivity 
and found that the gap ($\sim$ 700K) at low temperatures
is filled up at a temperature ($\sim$ 250K) lower than the gap size.
Therefore, they concluded that FeSi could not be understood with simple band 
models without correlation effect.
Low temperature properties were elucidated further,\cite{Hunt94,Paschen97}
and the importance of the correlation effects was underlined.\cite{Mandrus95}

The band calculation by Mattheis and Hamann\cite{Mattheiss93} gave the 
indirect gap of about 0.11eV, which is slightly
larger than the observed one.
Fu, et al.\cite{Fu94} also calculated the indirect gap of about 0.1eV.
The band width, however, was much larger than that assumed in the analysis of
ref.\cite{Jaccarino67}.

Recent {\it ab initio} band calculation with spin-orbit
interaction,\cite{Ohta94,Kulatov97} however, has given 
a rather small
gap size (indirect gap $\sim$ 0.02-0.03eV, direct gap $\sim$ 0.046-0.08eV).
It is well known that the local density approximation underestimates the
gap of insulators,\cite{Fulde91} and the precise estimate of the small gap size
is a difficult task.
Therefore, it seems reasonable to consider that the gap size obtained by the
band calculations {\it are} of the order of the observed value
and no 
large reduction of the gap due to correlation seems
necessary.

Some of the low temperature properties were calculated from the band
calculations\cite{Jarlborg95}, and the agreement with experiments were
obtained.
Temperature dependence of the optical conductivity was calculated
from the band calculations also,\cite{Fu94,Ohta94} but the experimentally
observed behavior\cite{Schlesinger93} was not reproduced.

On the other hand, high-resolution photoemission measurement was done by
Saitoh {\it et al.}\cite{Saitoh95}
They estimated the self-energy of the d-electrons from the difference 
between the photoemission spectra and the previous band 
calculation,\cite{Mattheiss93} and demonstrated a possible existence of the
strong renormalization of the electronic states at low energies.

In the above band calculations, minimum gap is given by the indirect gap, 
but it is close to a
direct gap, and the states near the Fermi level mainly consist of the 
d electrons.
Based on these, Fu and Doniach\cite{Fu95} proposed a two-band Hubbard 
model.
In a non-interacting case, it has a simple direct gap originating from a
hybridization between the two bands, and the Coulomb repulsion was assumed
to act only within each band.
The interband Coulomb repulsion and the exchange interaction were simply
neglected.
Therefore, this is the simplest model for FeSi, but may keep essential
features near
the Fermi level.
They studied many body effects using the perturbation theory with respect to 
the Coulomb interaction between d-electrons.
But their calculations seem to include some errors 
in the treatment of the proper symmetry of the self-energies.

The purpose of the present paper is to investigate their model correctly and 
to study the effects of correlation on the electronic structures and the
physical properties.
The correlation effects are not included in the spin fluctuation 
theory\cite{Takahashi79} at low temperatures.
We calculate the temperature dependence of the density of states, magnetic 
susceptibility, the specific heat and the dynamical conductivity using the 
self-consistent second order perturbation theory 
(SCSOPT)\cite{MullerHartmann89} together with the local
approximation, which becomes exact in the limit of large spatial
dimensions.\cite{Georges96}
Our basic formulation is the same as the previous one\cite{Fu95}, but the
obtained results are much different.
Our main conclusions are: (1) the gap is not much renormalized by the
correlation, and (2) the strong temperature dependence of the density of states
due to correlation results in the rapid collapse of the gap in the optical 
conductivity.
Although the model used here is a simplified one, these results can explain the experiments semi-quantitatively.
Similar calculations were performed on the periodic Anderson
model,\cite{Mutou94,Rozenberg96} and the rapid gap-filling was found.
Comparison of the present results with these calculations are also of interest.

\section{Model and Formulation}
The two-band Hubbard model introduced by Fu and Doniach\cite{Fu95}
 is written as
\begin{eqnarray}
H=
& &\sum_{ij\sigma}
t_{ij}(c_{i1\sigma}^\dagger c_{j1\sigma} -c_{2i\sigma}^\dagger 
c_{2j\sigma})\cr
&+&v\sum_{i\sigma}(c^\dagger_{i2\sigma}c_{i1\sigma}+
c^\dagger_{i1\sigma}c_{i2\sigma})\cr
&+&U\sum_{i\sigma}(n_{i1\uparrow}n_{i1\downarrow}+
n_{i2\uparrow}n_{i2\downarrow})\cr
&-&\frac{1}{2}g\mu_Bh\sum_{ia}(n_{ia\uparrow}-n_{ia\downarrow}),
\end{eqnarray}
where $c^\dagger_{ia\sigma}(c_{ia\sigma})$ creates (destroys) an electron
on site $i$ in band $a=1,2$ with spin $\sigma$.
The opposite signs in the hopping terms of the two bands provide a simple 
direct gap when they are hybridized via $v$.
The magnetic field is denoted by $h$.
Green's functions in the k-space can be
represented in the matrix form:
\begin{eqnarray}
\label{eq:last}
\!\!\!\!&{\bf G}_\sigma&\!({\bf k},\omega)
\equiv\pmatrix{
G^\sigma_{11}({\bf k},\omega)&G^\sigma_{12}({\bf k},\omega)\cr
G^\sigma_{21}({\bf k},\omega)&G^\sigma_{22}({\bf k},\omega)\cr
}\cr
& & \cr
& & \hspace{-8mm} =
\pmatrix{
\omega -\varepsilon_{\bf k}-\Sigma^\sigma_{11}({\bf k},\omega)
& -v-\Sigma^\sigma_{12}({\bf k},\omega) \cr
-v-\Sigma^\sigma_{12}({\bf k},\omega)
& \hspace{-3mm}\omega +\varepsilon_{\bf k}-\Sigma^\sigma_{22}
({\bf k},\omega) \cr
}^{-1}\!\!\!.
\end{eqnarray}
where $\varepsilon_{\bf k}$ denotes the Fourier transformation of $t_{ij}$. 
We choose $t_{ii}=0$.

To investigate the effects of electron correlation in
this model,
we adopt SCSOPT combined with the local 
approximation.\cite{MullerHartmann89}
The self-energy consists of the first and the second order terms, 
$\Sigma_{ab}^\sigma(\omega)= \Sigma_a^{(1)\sigma}\delta_{ab}+ 
\Sigma_{ab}^{(2)\sigma}(\omega)$, where 
$\Sigma_a^{(1)\sigma}=U(n_{a\uparrow}+n_{a\downarrow})/2
+U\sigma(n_{a\uparrow}-n_{a\downarrow})/2$, 
and $n_{a\sigma}$ denotes the number of electrons in 
band a with spin $\sigma$.
We implicitly introduce d-level energy $E_d=-U/2$ to keep particle-hole 
symmetry.  Then the first term $U(n_{a\uparrow}+n_{a\downarrow})/2$ in 
$\Sigma_a^{(1)\sigma}$ cancels with the $E_d$ term.
In the case of a finite magnetic field,
we must calculate $n_{a\sigma}$ self-consistently.

In SCSOPT, $\Sigma_{ab}^{(2)\sigma}(\omega)$ is calculated as
\begin{eqnarray}\label{eq:sigma}
\Sigma^{(2)\sigma}_{ab}(\omega)&=&
U^2 \int\!\!\!\int\!\!\!\int^\infty_{-\infty}
d\varepsilon_1 d\varepsilon_2 d\varepsilon_3
\rho^{-\sigma}_{ab}(\varepsilon_1)
\rho^{\sigma}_{ab}(\varepsilon_2)
\rho^{-\sigma}_{ab}(\varepsilon_3) \cr
\times & & 
\hspace{-5mm}\frac{f(-\varepsilon_1)f(\varepsilon_2)f(\varepsilon_3)
+f(\varepsilon_1)f(-\varepsilon_2)f(-\varepsilon_3)}
{\omega+\varepsilon_1-\varepsilon_2-\varepsilon_3+{\rm i}\delta},
\end{eqnarray}
where
\begin{equation}
\rho_{ab}^\sigma(\omega)=
-\frac{1}{\pi}{\rm Im}G_{ab}^{\sigma}(\omega+{\rm i}\delta)
\end{equation}
and
\begin{equation}
G_{ab}^\sigma(\omega)
=\frac{1}{N}\sum_{\bf k}G_{ab}^{\sigma}({\bf k},\omega).
\end{equation}
Here, $N$ is the number of sites.
These equations are converted into\cite{MullerHartmann89,Fu95}
\begin{eqnarray}
\Sigma^{(2)\sigma}_{ab}(\omega) &=&
-{\rm i}U^2\int^\infty_0 d\tau e^{{\rm i}\omega\tau}
\left[ \right.
B^{-\sigma}_{ab}(-\tau)A^\sigma_{ab}(\tau)
A^{-\sigma}_{ab}(\tau)\cr
& & \hspace{1.5cm}+A^{-\sigma}_{ab}(-\tau)B^\sigma_{ab}(\tau)
B^{-\sigma}_{ab}(\tau)
\left. \right],
\end{eqnarray}
where
\begin{eqnarray}
A^\sigma_{ab}(\tau)&=&\int^\infty_{-\infty}d\epsilon
e^{-{\rm i}\tau\varepsilon}\rho^\sigma_{ab}(\varepsilon)f(\varepsilon),\cr
B^\sigma_{ab}(\tau)&=&\int^\infty_{-\infty}d\epsilon
e^{-{\rm i}\tau\varepsilon}\rho^\sigma_{ab}(\varepsilon)f(-\varepsilon).
\end{eqnarray}
The basic formulation mentioned above is the same as in ref.\cite{Fu95}, but 
further analysis seems different from ours indicated below.

In the particle-hole symmetric case, each element of the Green's functions 
is calculated as follows:
\begin{equation}
G^\sigma_{11}(\omega)=G^\sigma_{22}(\omega)
=\frac{\omega-\Sigma^\sigma_{11}(\omega)}{\zeta_\sigma(\omega)}
F(\zeta_\sigma(\omega)),
\end{equation}
\begin{equation}
G^\sigma_{12}(\omega)=G^\sigma_{21}(\omega)
=\frac{v+\Sigma^\sigma_{12}(\omega)}{\zeta_\sigma(\omega)}
F(\zeta_\sigma(\omega)),
\end{equation}
\begin{equation}
\Sigma^\sigma_{11}(\omega)=\Sigma^\sigma_{22}(\omega)
\quad {\rm and} \quad
\Sigma^\sigma_{12}(\omega)=\Sigma^\sigma_{21}(\omega),
\end{equation}
where
\begin{eqnarray}
\hspace{-3mm}F(\zeta_\sigma(\omega))&\equiv&\frac{1}{N}\sum_{\bf k}
\frac{1}{\zeta_\sigma(\omega)-\varepsilon_{\bf k}},\cr
\zeta_\sigma(\omega)&\equiv&\sqrt{(\omega-\Sigma^\sigma_{11}(\omega))^2
-(v+\Sigma^\sigma_{12}(\omega))^2}.
\end{eqnarray}
To carry out ${\bf k}$-summation, we assume
the semi-circular shape of the density of states (
$D(\varepsilon)=(2/\pi W)\sqrt{1-\varepsilon^2/W^2}$) for the unperturbed 
band without hybridization.
Then we obtain
\begin{eqnarray}
F(z)
&=&\frac{2}{W^2}(z-\sqrt{z^2-W^2}).
\end{eqnarray}
Though Fu and Doniach assumed the Gaussian form to the density of states,
the difference in the shape of $D(\varepsilon)$
is not essential for the purpose of our calculation.

In the non-interacting case, the self-energy vanishes, so that we obtain the 
density of states as
\begin{eqnarray}\label{eq:rho0}
\rho^0_{11}(\omega)&=&\rho^0_{22}(\omega)\cr
&=& 
\frac{|\omega|}{\sqrt{\omega^2-v^2}}D(\sqrt{\omega^2-v^2})\theta(|\omega|-v)  
, \cr
\rho^0_{12}(\omega)&=&\rho^0_{21}(\omega)\cr
&=& \frac{v{\rm 
sgn}(\omega)}{\sqrt{\omega^2-v^2}}D(\sqrt{\omega^2-v^2})\theta(|\omega|-v). 
\end{eqnarray}
Note that $\rho^0_{12}(\omega)$ is an odd function.

\section{Calculation of Physical Quantities}
Starting from the equation of motion,\cite{Fetter71}
we obtain the following expression for the total energy
per site:
\begin{eqnarray}\label{eq:sh}
\epsilon&=&\frac{1}{2}\sum_\sigma\int^\infty_{-\infty}d\omega
f(\omega)\left[ \omega\{\rho^\sigma_{11}(\omega)+\rho^\sigma_{22}(\omega)\} 
\right.\cr
&&\qquad\qquad\qquad +v\{ 
\rho^\sigma_{12}(\omega)+\rho^\sigma_{21}(\omega)\}\cr
&& \hspace{-5mm} +\frac{1}{N}\sum_{\bf k}\left. \{\varepsilon_{\bf 
k}\rho^\sigma_{11}({\bf k},\omega)
+(-\varepsilon_{\bf k})\rho^\sigma_{22}({\bf k},\omega) \} \right],
\end{eqnarray}
where
\begin{equation}
\rho^\sigma_{ab}({\bf k},\omega)=-\frac{1}{\pi}{\rm Im}
G^\sigma_{ab}(\bf k,\omega+{\rm i}\delta).
\end{equation}
For $U=0$, the above equation reduces to the trivial form,
\begin{equation}
\epsilon_0=\sum_\sigma\int^\infty_{-\infty}d\omega
f(\omega)\omega\left[\rho^\sigma_{11}(\omega)+\rho^\sigma_{22}(\omega)\right].
\end{equation}
The ${\bf k}$-sums in the brackets in eq.(\ref{eq:sh}) are calculated as 
follows:
\begin{eqnarray}
\frac{1}{N}\sum_{\bf k}\varepsilon_{\bf k}
G^\sigma_{11}({\bf k},\omega)
&=&\frac{1}{N}\sum_{\bf k}(-\varepsilon_{\bf k})
G^\sigma_{22}({\bf k},\omega)\cr
&=&-1+\zeta_\sigma(\omega)F(\zeta_\sigma(\omega)).
\end{eqnarray}
The specific heat is given by
\begin{equation}
C_{\rm V}=\left(
\frac{\partial \epsilon}{\partial T}
\right)_{\rm V}.
\end{equation}
This differentiation is carried out numerically.

The magnetic moment per site is given by
\begin{eqnarray}
m(T,h)=\frac{g\mu_B}{2}\int^\infty_{-\infty}d\omega f(\omega)
\!\!\!\!\!\!
&&\left[\right.\rho^\uparrow_{11}(\omega)-\rho^\downarrow_{11}(\omega)\cr
&+&\rho^\uparrow_{22}(\omega)-\rho^\downarrow_{22}(\omega)\left.\right].
\end{eqnarray}
The magnetic susceptibility is calculated by $\chi(T)=m(T,h)/h$ at
sufficiently small $h$.
Furthermore,
using the fact that the vertex correction in
the dynamical conductivity drops out in $d=\infty$ theory,\cite{Georges96} 
we also obtain the expression for the dynamical conductivity
in Fu-Doniach model as
\begin{eqnarray}
{\rm Re}
\sigma(\omega,T)&=&2\pi\sum_\sigma
\int^\infty_{-\infty}d\varepsilon d\nu D(\varepsilon)
\frac{f(\nu)-f(\nu+\omega)}{\omega}\cr
&& \hspace{-2cm}\times
[\rho^\sigma_{11}(\varepsilon,\nu)\rho^\sigma_{11}(\varepsilon,\nu+\omega)
-\rho^\sigma_{12}(\varepsilon,\nu)\rho^\sigma_{12}(\varepsilon,\nu+\omega)
],
\end{eqnarray}
where
\begin{eqnarray}
\rho^\sigma_{ab}(\varepsilon,\omega)
=\frac{1}{N}\sum_{\bf k}\delta(\varepsilon_{\bf k}-\varepsilon)
\rho^\sigma_{ab}({\bf k},\omega).
\end{eqnarray}
For $U=0,\ h=0$ this equation becomes
\begin{eqnarray}
\lefteqn{\pi\rho_{11}(\omega/2)\frac{v^2}{(\omega/2)^3}
(1-2f(\omega/2))}\cr
&& \hspace{-5mm} +4\pi\delta(\omega)\int^\infty_{-\infty} d\nu 
D(\sqrt{\nu^2-v^2})\sqrt{\nu^2-v^2}
\frac{-f^\prime(\nu)}{|\nu|}.
\end{eqnarray}
The first and second terms describe the interband and intraband (Drude) 
contributions, respectively.

\section{Results}
We take the unit of energy as $W=1$ and choose $v=0.125,$
which yields the gap of $2v=0.25$ for $U=0$.

In Figs. 1(a) and 1(b), we show the self-energies.
Though the diagonal parts have rather similar shapes to
those of the single-band Hubbard model,
the off-diagonal parts have the opposite parity compared to the diagonal 
one.
This can be understood from the odd parity of $\rho^0_{12}(\omega)$ in 
eq.(\ref{eq:rho0}), since Im$\Sigma_{ab}^\sigma$ has the same parity as 
$\rho_{ab}(\omega)$ (see eq.(\ref{eq:sigma})).
In ref.\cite{Fu95}, both Im$\Sigma_{11}$ and Im$\Sigma_{12}$ are treated as 
even functions, which implies that both Re$\Sigma_{11}$ and Re$\Sigma_{12}$ 
have odd parity and vanish at $\varepsilon=0$ in their calculation.
In our proper treatment, ${\rm Re}\Sigma_{12}$ is an even function and has a 
positive value at low energies and is added to the mixing $v$ in $G_{ab},$ 
so that the reduction of the gap due to the renormalization factor
$z=(1-\partial\Sigma_{aa}(\omega)/\partial\omega|_{\omega=0})^{-1}$ is 
almost compensated by this
increase of the effective mixing (see below).

\begin{figure}\vspace{0.5cm}
\epsfysize=6cm
\centerline{\epsfbox{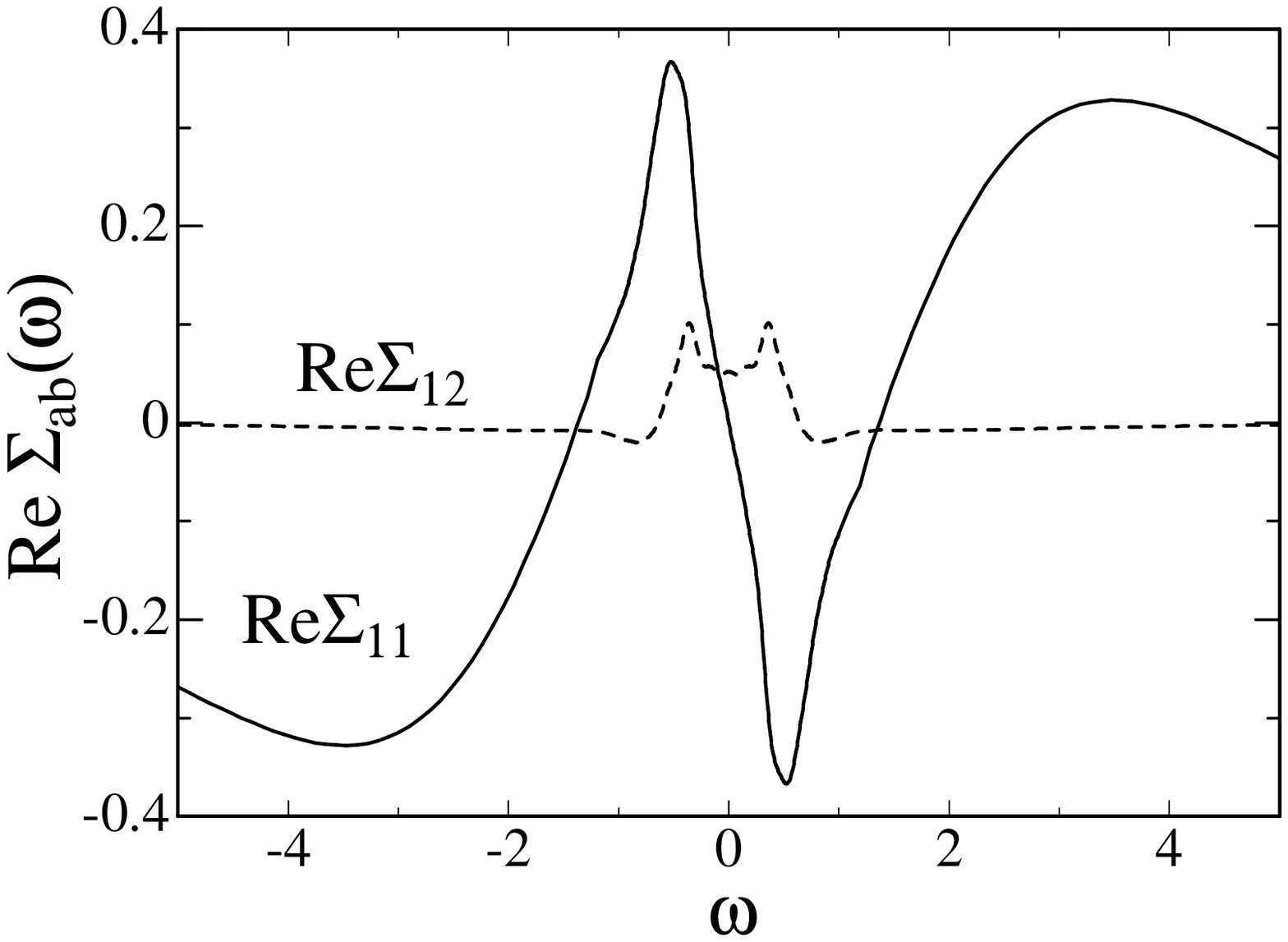}}
\epsfysize=6cm
\centerline{\epsfbox{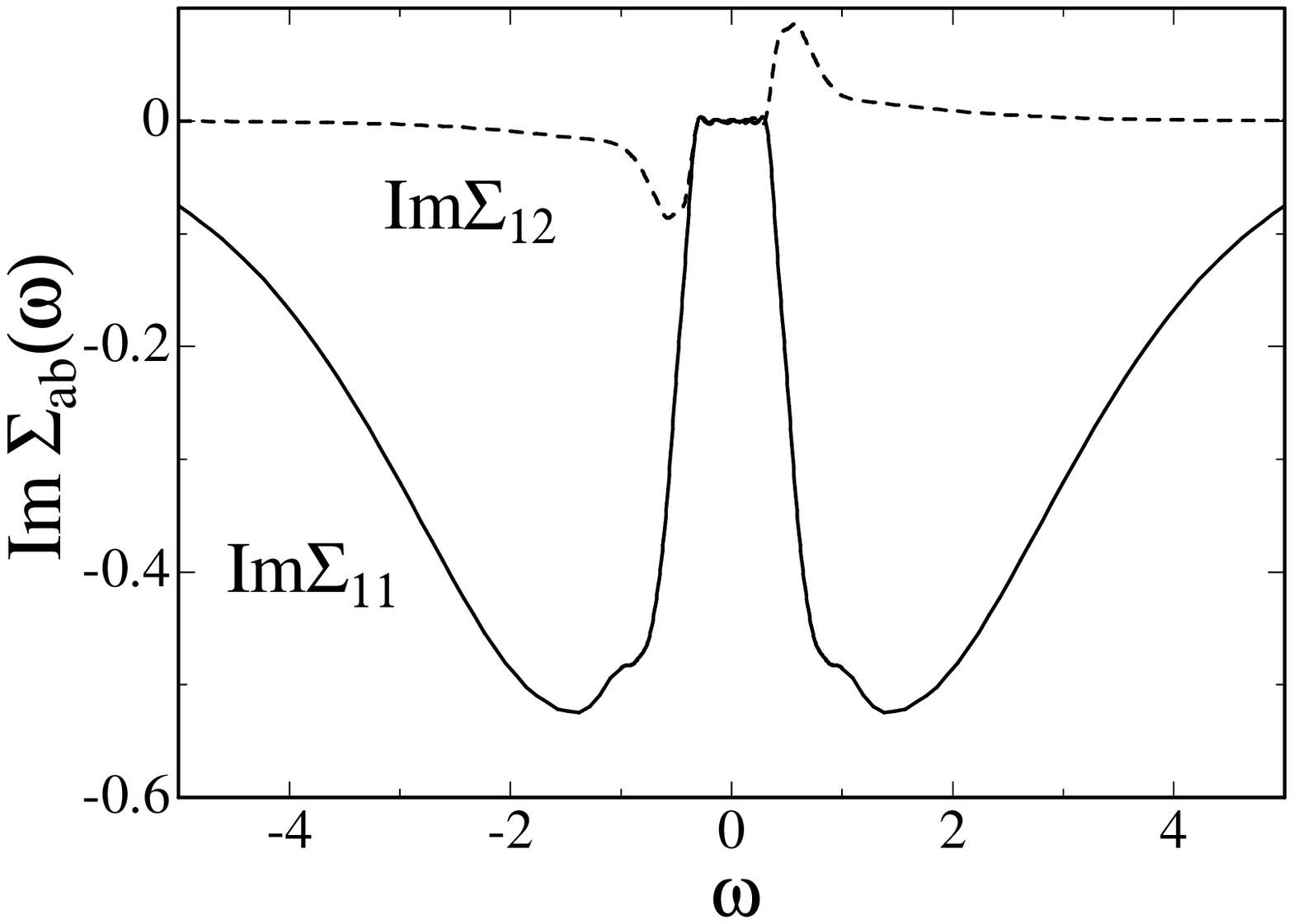}}
\caption{The real and imaginary parts of the self-energies 
$\Sigma_{ab}(\omega)$ are shown for $U=2$, $v=0.125$ and $T=0$.  The full 
and the broken lines indicate the diagonal ($\Sigma_{11}$) and the 
off-diagonal ($\Sigma_{12}$) parts, respectively.}
\label{fig:1}
\end{figure}

\begin{figure}
\epsfysize=6cm
\centerline{\epsfbox{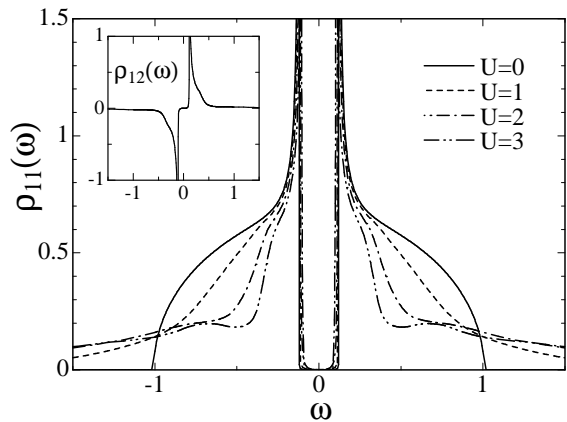}}
\epsfysize=6cm
\centerline{\epsfbox{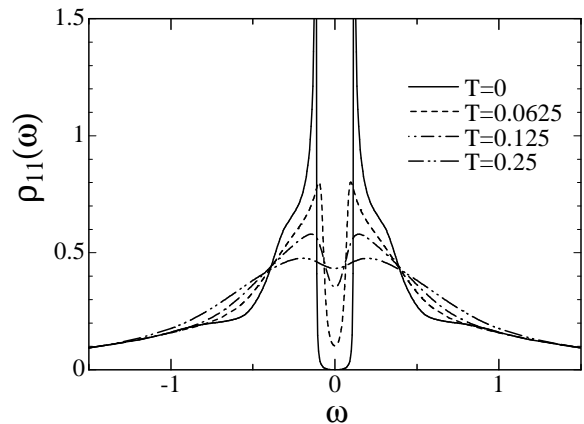}}
\caption{(a) The quasi-particle density of states $\rho_{11}(\omega)$ at 
$T=0$ is shown for $U=$0, 1, 2 and 3.  The insert indicates 
$\rho_{12}(\omega)$ for $T=0$ and $U=2$. (b) $\rho_{11}(\omega)$ for $U=2$ 
at various temperatures are shown.}
\label{fig:2}
\end{figure}

In Fig. 2(a) and 2(b), the quasi-particle density of states obtained with
the above self-energies are shown.
In the present model, the density of states for $U=0$ has a gap of width 
$2v$
that has sharp edges and diverges at $|\omega|=v.$
So we take $\delta$ finite ($\delta=10^{-4}$ except in $\sigma(\omega)$)
 in $\omega +{\rm i}\delta$
in order to make numerical calculation easy.
It is seen in Fig.2(a) that the gap size is not significantly reduced when 
$U$ increases.
The reason is mentioned above.
This result is in strong contrast with the periodic Anderson model, in
which the gap is strongly reduced by the Coulomb repulsion between
f-electrons.\cite{Mutou94,Saso97}

We have also investigated a variation of the present model, where the band 
energies $\varepsilon_{\bf k}$ and $-\varepsilon_{\bf k}$ in the band 1 and 
2 are replaced with $\varepsilon_{\bf k}$ and $-b\varepsilon_{\bf k}$, 
respectively, and the Coulomb repulsion $U$ in band 1 is replaced with 
$bU$.
If we set $b=0$, the band 2 becomes dispersionless and the band b becomes 
free, so we obtain the periodic Anderson model, in which the gap is known to 
be reduced due to the correlation.
We confirmed that the gap is continuously reduced when we change from $b=1$ 
 to 0.

The density of states varies with temperature when $U$ is finite.
In Fig.2(b), we show the density of states for $U=2$ at various 
temperatures.
The gap is almost filled up at the temperature of the order of the gap.
It is filled up faster for larger $U$.

Figure 3 shows the temperature dependence of
the electronic specific heat per site.
The peak in the $U=0$ curve splits into lower and higher
temperature regions due to correlation effect at finite $U$.
This tendency becomes less remarkable if we use the density of states 
at $T=0$ to calculate $C_V(T)$ at finite temperatures. 
It indicates the importance of the temperature dependence of the 
density of states itself in the calculation of the physical quantities. 

\begin{figure}
\epsfysize=6cm
\centerline{\epsfbox{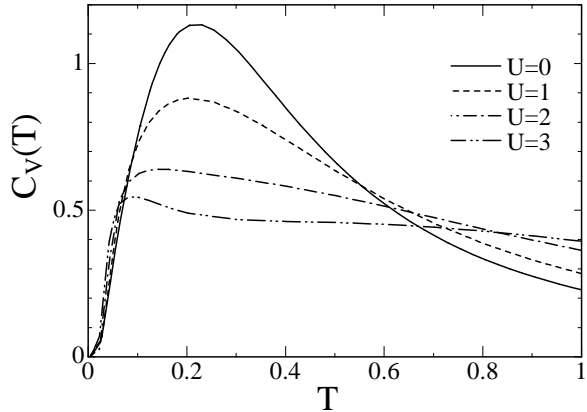}}
\caption{Temperature dependence of the specific heat is displayed for $U=$0, 
1, 2 and 3.}
\label{fig:3}
\end{figure}
\begin{figure}
\epsfysize=6cm
\centerline{\epsfbox{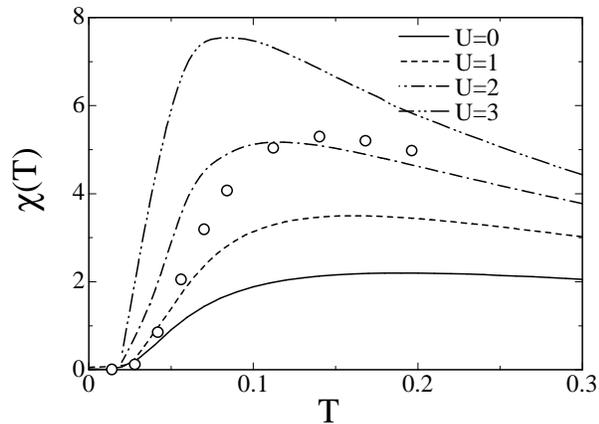}}
\caption{Temperature dependence of the spin susceptibility in unit of 
$(g\mu_B)^2/W$ is displayed for $U=$0, 1, 2 and 3.  The circles indicate the 
experimental result[1]  
(see the text in V).}  
\label{fig:4}
\end{figure}
The temperature dependence of the spin susceptibility per site for different 
magnitude of $U$ are displayed in Fig. 4.
As $U$ increases, $\chi(T)$ is largely enhanced in the intermediate 
temperature region, and the peak positions shift to lower temperatures,
similar to the specific heat.

In Fig. 5, the dynamical conductivity for $U=2$ at various temperatures is 
shown.
At $T=0$, $\sigma(\omega)$ has a gap of about 0.23, although the gap edge is 
rounded due to the finite $\delta=2\times 10^{-3}$.
The sharp peak at the gap edge reflects the peaks in the density of 
states.
$\sigma(\omega)$ varies with temperature faster than
the density of states because it consists of the convolution of
the two spectral functions and the Fermi distribution function.
As a result, the gap is filled up almost completely at the temperature
which is about a third of the gap size.
If we take a smaller value of $U$, e.g. $U=1$, the gap is filled up more 
slowly, at a temperature which is a half of the gap.
\begin{figure}
\epsfysize=6cm
\centerline{\epsfbox{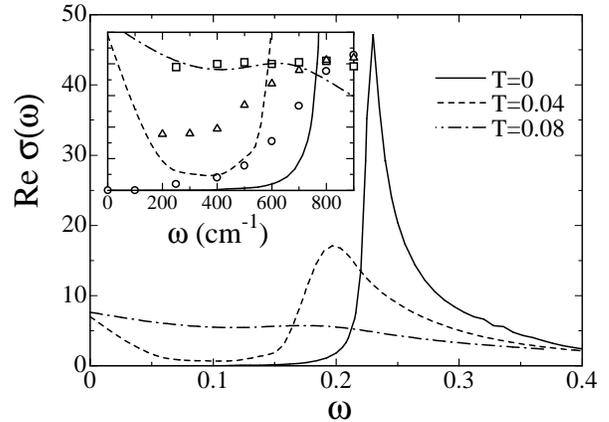}}
\caption{The dynamical conductivity for $U=2$ is shown for 
$T$=0, 0.04 and 0.08.
The insert shows the comparison of the calculation with the 
experiments at $T$=20K(circles), 150K(triangles) and 
250K(squares).[8]
}
\label{fig:5}
\end{figure}

\section{Discussions}
Although the model used in the present study is oversimplified as a model for FeSi, we try to compare the obtained results with the experiments.
The gap size of FeSi is estimated from the activation energy of the magnetic  
susceptibility and the resistivity.
Both yield the gap size of about 650-700K.\cite{Schlesinger93}
In the measurement of $\sigma(\omega)$ at $T=$20K, there is a threshold at
about 500cm$^{-1} \sim$ 700K.\cite{Schlesinger93}
Therefore, we first take 700K as the gap at $T=0$.

If we regard the gap size 0.23 for $U=2$ in $\sigma(\omega)$ to be 
corresponding to the gap of 700K, the peak position in $\chi(T)$ ($T \sim$0.11)
corresponds to about 340K,
which is too low compared the experiment (500K\cite{Jaccarino67}).
If we take $U=1$ with the gap $\sim$ 0.24 in $\sigma(\omega)$, the peak 
position in $\chi(T)$ becomes $\sim$ 0.11 $\sim$ 470K,
which is closer to the observed one.
Recent experiment at a lower temperature ($T=$4K),\cite{Damascelli97} 
however, reported a little larger gap size $\sim$ 570cm$^{-1} \sim$ 820K.
Use of this value results in the peak position of $\chi(T)$ with $U=2$ at 
400K, which is not much different from the experiment 
(see the circles in Fig.4). 
The absolute value is fitted {by adjusting the value
of the effective g-factor ($g=$2.53), which represents the effects of the
multi-band 
and the electron number in the real system.
}

As we have shown in the previous section, the density of states in Kondo 
insulators depend on temperature.
As a result, the activation energy analysis from $\chi(T)$ or resistivity 
may underestimate the gap size at $T=0$. 
We therefore conclude that $U=2$ is more appropriate 
to fit both $\sigma(\omega,T)$ 
and $\chi(T)$ within the present model, although refinement of the model and 
the calculation method is definitely needed.

There is a large ambiguity in the experimental determination of the 
electronic part of the specific heat,\cite{Jaccarino67} 
so that we did not compare the peak position of 
$C_{\rm V}(T)$ with experiments.

In the present paper, we carefully investigated the two-band Hubbard model 
proposed by Fu and Doniach as a simplest model for FeSi.
The electron correlation was taken into account by use of the 
self-consistent second-order perturbation theory.
We calculated the temperature dependence of spin susceptibility, electronic 
specific heat and the dynamical conductivity.
The observed temperature dependence of $\chi(T)$ and $\sigma(\omega,T)$ were 
semi-quantitatively explained by the present model if we fit the optical gap 
size to the recent experimental one and choose $U/W=2$.
However, $\sigma(\omega)$ in the present calculation has a sharp peak at the 
gap edge, which is not observed in experiments.
This peak originates from the large peaks in the density of states, and would be
suppressed if we use the density of states obtained from band calculations\cite{Mattheiss93}-\cite{Kulatov97}, or if we introduce disorder\cite{Rozenberg96}.
Secondly, the filling up of the gap of the measured $\sigma(\omega)$ with 
increasing temperature does not seem to be compensated by the sufficient 
decrease in the region right above the gap.
A possible solution for the case of the periodic Anderson model was discussed in \cite{Rozenberg96}.
Thirdly, the observed $\sigma(\omega)$ has a large tail 
in the high frequency region, 
irrespective of temperature.
To consider these features correctly, one needs to start from a more 
realistic energy band model.
We like to perform such calculations in our next studies.

\section*{Acknowledgements}
One of the authors (T.S.) thanks to Professor H. Yamada for informing them the
paper on recent band calculation\cite{Kulatov97}.

\end{document}